\documentclass[10pt]{article}
\usepackage[english]{babel}
\usepackage{graphicx}    

\begin{document}
\begin{center}
{Sterile Neutrino as Dark Matter candidate from CMB alone}

\vspace{0.5cm}
L.A. Popa and A. Vasile

\vspace{0.2cm}
Institute for Space Sciences,\\
Bucharest - Magurele, Ro-077125 Romania
\end{center} 
\begin{abstract}
Distortions of CMB temperature and polarization maps caused by gravitational lensing, 
observable with high angular resolution and sensitivity, can be used to constrain the 
sterile neutrino mass, $m_s$, from CMB data alone.
We forecast $m_s>$1.75 keV from {\sc Planck} and $m_s>$4.97 keV 
from {\sc Inflation Probe} at 95\% CL, by using the CMB weak lensing extraction. 
\end{abstract}

\section{Introduction}\label{intro}
The confluence of the most recent experimental data of the cosmic
microwave background (CMB) anisotropies, large-scale structure (LSS)
galaxy surveys, supernovae luminosity distance, Lyman-$\alpha$
forest and Hubble parameter, have lead in specifying the $\Lambda$CDM model 
as the cosmological concordance model \cite{Spergel06}. According to this model 
the Universe is spatially nearly flat with energy densities 
of $\Omega_{CDM}$=0.27$\pm$0.07 in cold dark matter (CDM) particles, 
$\Omega_b$=0.044$\pm$0.004 in baryons, $\Omega_{\Lambda}$=0.70$\pm$0.03 
in dark energy and Hubble constant $H_0$=72$\pm$5 km s$^{-1}$ Mpc$^{-1}$.

The direct confirmation of this theory was the detection of the acoustic Doppler peaks
structure of the CMB angular power spectrum. Further successes are related to the correct
prediction of the hierarchical structure formation via gravitational instability, 
the abundance of clusters at small redshifts, the spatial distribution and the number 
density of galaxies, the LSS matter power spectrum, the Lyman-$\alpha$  forest 
amplitude and spectrum.   

Despite its successes on large scales, the $\Lambda$CDM model produces too  much
power on small scales.
In general, the observed structures have softer cores, 
lower concentrations and are less clumped than those predicted by the
$\Lambda$CDM model \cite{Primack98}. 

A possibility to alleviate the accumulating contradiction between  $\Lambda$CDM 
predictions on small scales and observations is to add properties to the dark matter sector, 
relaxing the hypothesis on dark matter as being cold. 
Free streaming due to thermal motion  of dark matter particles  is the simplest known mechanism for smearing out small scale structure. 
For example, the velocity dispersion of warm dark matter (WDM) particles 
is sufficient to alleviate some of these problems \cite{Padm95}. 

Sterile neutrinos are considered the most promising WDM candidates.
Lower limits on sterile neutrino mass have been placed from various observational probes\footnote{Throughout the paper the sterile neutrino mass is quoted at 95\% CL.}. \\  
The combination of CMB measurements, LSS and Lyman-$\alpha$ forest 
power spectra lead to  $m_s > 1.7$ keV
 with a further 
improvement to $m_s > 3$ keV  when  
high-resolution Lyman-$\alpha$ spectra  are considered \cite{Aba06}.\\ 
The upper limit of sterile neutrino mass is constrained by 
the limits on its radiative decay from Virgo cluster observations \cite{Aba01}
and by the observations of the diffuse X-ray background \cite{Boy05}. 
The combination of all above constraints allows the range 
1.7 keV$< m_s <$8.2 keV for the mass of sterile 
neutrino as dark matter candidate. 

A significantly more stringent lower limit constraint, $m_s > 14$ keV, was placed by using 
the Lyman-$\alpha$ forest power spectrum and high-resolution spectroscopy observations in combination with CMB and galaxy clustering data, excluding sterile neutrino  as dark matter candidate \cite{Sel06}.

\section{Sterile neutrino mass from CMB lensing extraction}

Like active neutrinos, sterile neutrinos can not cluster via gravitational instability 
on scales below the free-streaming scale,  
with  important implications for the  growth of density perturbations at late times.  

The alteration of the gravitational potential changes 
the gravitational lensing of the CMB photons. 
Week lensing introduces a deflection field so that the deflection 
angle power spectrum $C^{dd}_l$ and the projected gravitational potential power spectrum $C^{\Phi\Phi}_l$
are related through $C^{dd}=l(l+1)C^{\Phi\Phi}_l$.

We modified the CMB anisotropy code CAMB \cite{Lew00} to compute the  lensed  CMB temperature and 
polarization anisotropy power spectra and  the projected gravitational potential power spectrum in the presence of a sterile neutrino component. 
We include in the computation the momentum-dependent sterile neutrino 
phase-space distribution function \cite{Aba01},  
its unperturbed and perturbed energy density and pressure, 
energy flux and shear stress. 
 
The CMB weak lensing map can be reconstructed from the statistical analysis of the CMB temperature and polarization anisotropy maps.\\
To evaluate the ability of the future CMB experiments as {\sc 
Planck} \cite{Blue05} 
and the hypothetical {\sc Inflation Probe}  to detect the 
sterile 
neutrino mass 
we employ the quadratic estimator method \cite{Pero06} to compute the expected noise power spectrum from lensing extraction. \\
The experimental parameters for the CMB projects considered in this work are presented in Table~1. For each experiment we construct the covariance matrix:
\begin{eqnarray}
{\bf C}=\left(\begin{array}{ccc}
C^{TT}_l+N^{TT}_l  &     C^{TE}_l       &     C^{Td}_l \\
       C^{TE}_l    & C^{EE}_l+N^{EE}_l  &      0    \\
       C^{Td}_l    &        0           & C^{dd}_l+N^{dd}_l 
\end{array}\right) \nonumber
\end{eqnarray}
where $C^{XX}_l$ with $X=\{ T,E\}$ are the power spectra of primary anisotropies,
and $N^{XX}_l$ are  the corresponding noise power spectra, $C^{dd}_l$ is deflection angle power 
spectrum, $N^{dd}_l$ the noise power spectrum associated to the lensing extraction and 
$C^{Td}_l$ is the power spectrum of the cross-correlation between the temperature and deflection angle. 

In the left panel of Figure~1 we present  the above power spectra
obtained for our {\it fiducial} model, 
the $\Lambda$CDM  concordance model. 
We assume adiabatic initial conditions, primordial scalar density perturbations with scalar spectral index $n_s$=0.95 and three active neutrino flavors with a total mass $m_{\nu}=$0.7eV.
\begin{figure}
\begin{center}
\includegraphics[height=8cm,width=13cm]{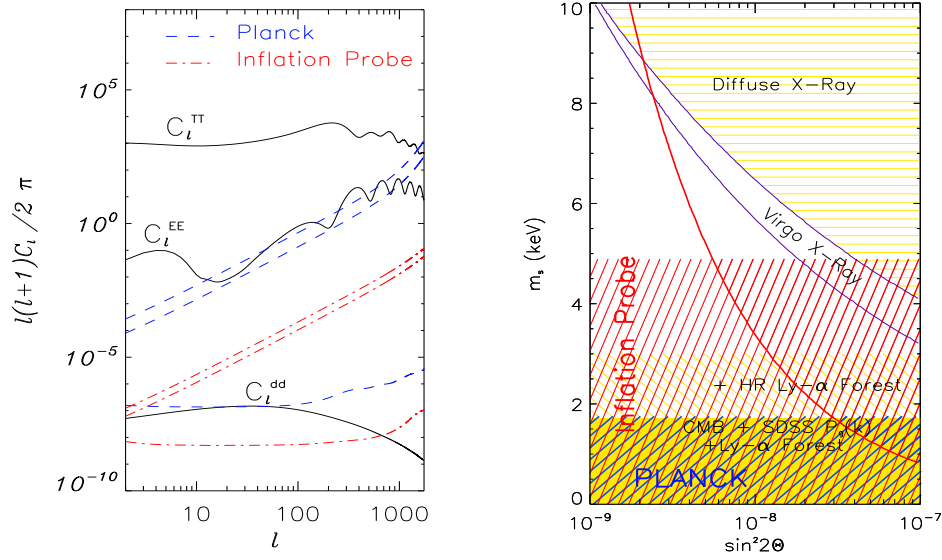}
\caption{{\bf Left:} The CMB temperature and polarization power spectra, 
$C^{TT}_l$ and  $C^{EE}_l$, and the deflection angle power spectrum, $C^{dd}_l$, for the {\it fiducial} model. Dashed lines (from top to bottom) are the noise power spectra 
associated to temperature, polarization and lensing extraction for {\sc Planck} (blue dashed)  and {\sc Inflation Probe} (red dash-dotted).
{\bf Right:} Constraints on sterile neutrino parameter space potentially obtained 
by the future  {\sc Planck} and {\sc Inflation Probe} 
from CMB weak lensing extraction compared with other experimental constraints (see also the text). 
The red line reprezents the constraint on sterile neutrino parameter space from an ideal noise-less experiment.}
\label{aba:fig1}
\end{center}
\end{figure}

The right panel of Figure~1  presents the constraints on sterile neutrino 
parameter space that can be potentially obtained 
by the future experiments {\sc Planck} and {\sc Inflation Probe}
from the CMB weak lensing extraction. \\ 

 \section{Conclusions}
 
In this paper we have studied the ability of the future CMB projects 
to constrain the sterile neutrino as dark matter candidate by using the CMB weak 
lensing extraction. Weak lensing offers several advantages: it probes a larger range of scales,
unlike the Lyman-$\alpha$ forest data and does not involve any light-to-mass bias, 
unlike the galaxy redshift surveys data. 
  
We found  $m_s>$1.75 keV from {\sc Planck} and 
$m_s >$4.97  keV from {\sc Inflation Probe} at 95\% CL 
for sterile neutrino mass from CMB data alone. 
\begin{table}      
\caption{Experimental characteristics of the CMB experiments considered in this work:
$\nu$ is the frequency of the channel, $\theta_b$ is the FWHM, 
$\sigma_T$ and $\sigma_P$ are the sensitivities per pixel for temperature and polarization
maps.}

\vspace{0.3cm}
\begin{center}
\begin{tabular}{ccccc}
\hline
Experiment & Frequency(GHz) & FWHM (arc-minutes) & $\sigma_T (\mu K) $ & $\sigma_P (\mu K)$ \\
\hline
              & 100 & 9.5 &6.8 & 10.9 \\
{\sc Planck}  & 143 & 7.1 &6.0 & 11.4 \\
	      & 217 & 5.0& 13.1 & 26.7 \\
\hline
                & 70& 6.0 & 0.29& 0.41 \\
{\sc Inflation Probe }                & 100&4.2 & 0.42 & 0.59 \\
		& 150 & 2.8 & 0.63 & 0.88 \\
		& 220 & 1.9 & 0.92 & 1.30 \\
\hline
\end{tabular}
\end{center} 
\end{table}

\end{document}